\documentclass[a4paper,12pt]{article}
\usepackage[english]{babel}
\usepackage{amsmath,graphicx}

\begin{document}
%
\title{A fast algorithm of coexisting phases compositions calculation in binary systems}
\author{A.Yu.~Zakharov, A.A.~Schneider \\  \\
Yaroslav-the-Wise Novgorod State University\\ 
Novgorod the Great, 173003, RUSSIA\\
e-mail: anatoly.zakharov@novsu.ru }
\date{}

\maketitle

\begin{abstract}
A simple fast algorithm of the conodes calculation in binary systems is proposed. The method is based on exact solution of the problem on common tangent to pair of approximating parabolas. Sequence of approximating parabolas pairs having second order tangency points on separated concave parts of free energy isotherm is generated. The sequences of the tangency points pairs conodes with approximating parabolas converge to equilibrium compositions of phases in binary systems.\\
\end{abstract}


\section {Introduction}
\begin{figure}
\includegraphics[width=5.0in]{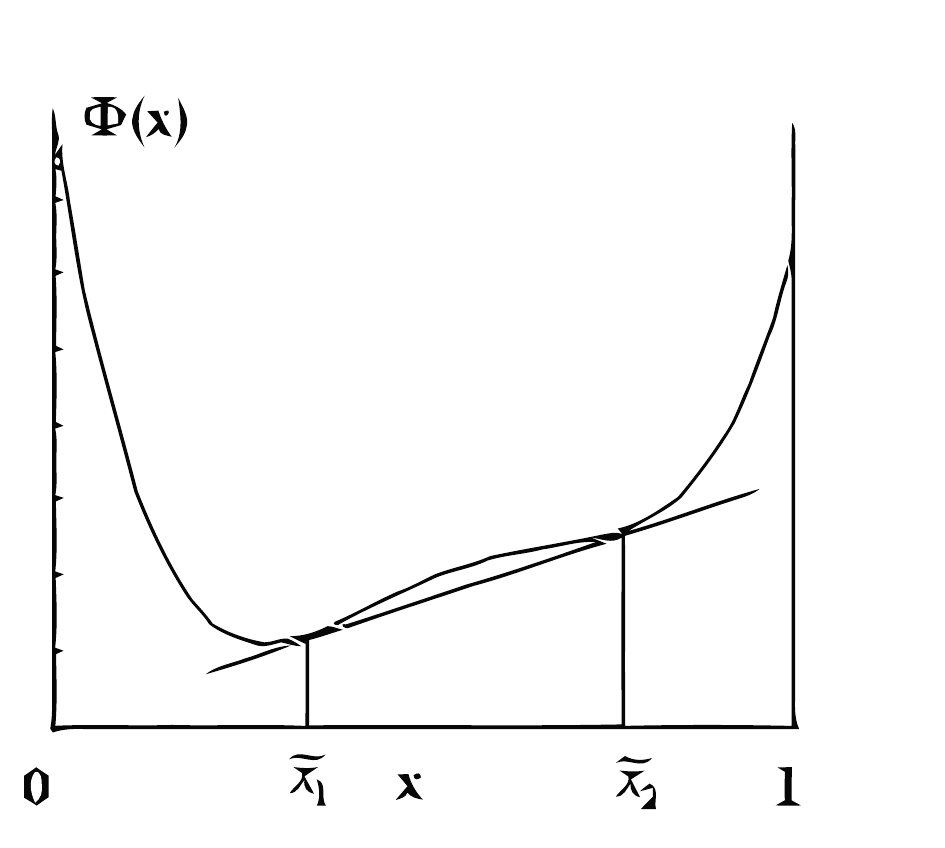} 
\caption{Schematic free energy $\Phi(x)$~---~composition $x$ diagram for a binary system in two-phase region.}
\label{fig:Helm}
\end{figure}

Free energy $\Phi(x)$ of a binary system at fixed volume $V$ or pressure $P$ and temperature $T$ depends on the system composition $x$ only.  A typical form of function $\Phi(x)$ graph in two-phase region presented on Figure~\ref{fig:Helm}. Equilibrium compositions $\tilde{x}_1$, $\tilde{x}_2$ of the coexisting phases in binary systems by known free energy $\Phi(x)$ are determined by tangency points of the conode\footnote{conode is the common tangent to separated concave parts of function $\Phi(x)$ \cite[p.~71]{Hillert}} with function~$\Phi(x)$ graph.

Thus, the problem of $\tilde{x}_1$ and $\tilde{x}_2$ evaluation reduces to pure mathematical task of the conode for given function~$\Phi(x)$ finding. There are several methods of this task solutions~\cite{Hillert2}. Most of them require a number iterations to have accuracy of the order of $10^{-3}$. 

The aim of this paper is development of an algorithm for evaluation of the equilibrium compositions of phases in binary systems with same accuracy in 2--3 iterations.

\section{The algorithm}

The algorithm of the problem solution consists the following steps
\begin{enumerate}
    \item Choice in the {\em zero-order approximation} some arbitrary points $x_1^{(0)}$, $x_2^{(0)}$ in first and second concavity intervals, respectively. These points are the second order point of tangency with tangent parabolas.
    \item Derivations of the equations of tangent parabolas to curve $\Phi(x)$ in points $x_1^{(0)}$, $x_2^{(0)}$, respectively.
    \item Exact analytical finding of the conode to approximating parabolas and determination the point of contact conode with approximating parabolas.
    \item Choice in the {\em first-order approximation} $x$-coordinates $x_1^{(1)}$, $x_2^{(1)}$ of the point of contact conode with approximating parabolas and transfer to point. 
\end{enumerate}
Performance of iterations stops after achievement of the set accuracy.

\section{Analytical calculations}

In zero approach we shall choose points on the different parties of a convex site of a curve $ \Phi (x) $. The equations of parabolas approximating free energy in vicinities of points $x_i^{(0)},~(i=1,2)$ have the following form
\begin{eqnarray}\label{par-12}
f_1^{(0)}\left(x \right) & = & a_1^{(0)} + b_1^{(0)} \left(x - x_1^{(0)} \right) + \frac{c_1^{(0)}}{2} \left(x - x_1^{(0)} \right)^2; \label{par1}\\
f_2^{(0)}\left( x\right) & = & a_2^{(0)} + b_2^{(0)} \left(x - x_2^{(0)} \right) + \frac{c_2^{(0)}}{2} \left(x - x_2^{(0)} \right)^2,\label{par2}
\end{eqnarray}
where
\begin{equation}\label{abc}
    a_k^{(0)} = \Phi\left(x_k^{(0)} \right); \quad  b_k^{(0)} = \Phi'\left(x_k^{(0)} \right); \quad c_k^{(0)} = \Phi''\left(x_k^{(0)} \right)
\end{equation}
are values of function $\Phi\left(x \right)$ and its derivatives of the first and second orders in points~$x_1^{(0)}$ and $x_2^{(0)}$.
The equations of tangents to each of parabolas in corresponding points~$x_1^{(1)}$ and $x_2^{(1)}$ have the following form:
\begin{eqnarray}
y_1^{(1)}\left( x \right) & = & f_1^{(0)} \left(x_1^{(1)} \right)  + f_1^{(0)'} \left(x_1^{(1)} \right)\left(x - x_1^{(1)} \right); \label{tan1}\\
y_2^{(1)}\left( x \right) & = & f_2^{(0)} \left(x_2^{(1)} \right)  + f_2^{(0)'} \left(x_2^{(1)} \right)\left(x - x_2^{(1)} \right) \label{tan2}.
\end{eqnarray}

The condition of coincidence of these tangents corresponds to system of two equations concerning two abscissas points of a contact $x_1^{(1)} $ and $x_2^{(1)} $ to the general tangent to both approximating parabolas (\ref {par1}) and (\ref {par2}):
\begin{equation}\label{iter-1}
\left\{
\begin{array}{l}
{\displaystyle  b_1^{(0)}  + c_1^{(0)} \left(x_1^{(1)} - x_1^{(0)} \right) = b_2^{(0)}  + c_2^{(0)} \left(x_2^{(1)} - x_2^{(0)} \right);  }\\
{\displaystyle a_1^{(0)} + b_1^{(0)} \left(x_1^{(1)} - x_1^{(0)} \right) + \frac{c_1^{(0)}}{2} \left(x_1^{(1)} - x_1^{(0)} \right)^2 - x_1^{(1)} \left[ b_1^{(0)}  + c_1^{(0)} \left(x_1^{(1)} - x_1^{(0)} \right) \right] }  \\ \\
{\displaystyle =  a_2^{(0)} + b_2^{(0)} \left(x_2^{(1)} - x_2^{(0)} \right) + \frac{c_2^{(0)}}{2} \left(x_2^{(1)} - x_2^{(0)} \right)^2 - x_2^{(1)} \left[ b_2^{(0)}  + c_2^{(0)} \left(x_2^{(1)} - x_2^{(0)} \right) \right]   }.
\end{array}
\right.
\end{equation}

The solution of system (\ref {iter-1}) in terms of a deviation from initial approximation 
$ \Delta x_i ^ {(1)} =x_i ^ {(1)}-x_i ^ {(0)} $ leads to following result 
\begin{equation}
\label{G1}
\Delta x_2^{(1)} =\frac{1}{c_2^{(0)}}\left[b_1^{(0)}-b_2^{(0)}+c_1^{(0)}\Delta x_1^{(1)}\right],
\end{equation}
and $\Delta x_1^{(1)}$ satisfies to the quadratic equation
\begin{equation}
A^{(1)}(\Delta x_1^{(1)})^2+B^{(1)}\Delta x_1^{(1)}+C^{(1)}=0,
\label{G2}
\end{equation}
with following coefficients
\begin{equation}\label{A1}
A^{(1)}=\frac{(c_1^{(0)}-c_2^{(0)})}{2},
\end{equation}
\begin{equation}\label{B1}
B^{(1)}=c_2^{(0)}\left(x_2^{(0)}-x_1^{(0)}\right)- \left(b_2^{(0)}-b_1^{(0)}\right),
\end{equation}
\begin{equation}
\label{C1}
C^{(1)}=\frac{c_2^{(0)}}{c_1^{(0)}} \left(b_1^{(0)}(x_2^{(0)}-x_1^{(0)})-(a_2^{(0)}-a_1^{(0)}) \right)+\frac{(b_2^{(0)}-b_1^{(0)})^2}{2 c_1^{(0)}}.
\end{equation} 

It is necessary to choose from two solutions of the equation~(\ref{G2}) the solution that has a finite limit at $(c_1 - c_2) \to 0$:
\begin{equation}
\Delta x_1^{(1)}=\frac{1}{2A^{(1)}}\left[-B^{(1)}+\sqrt{(B^{(1)})^2-4A^{(1)}C^{(1)}}\right].
\label{z11}
\end{equation}

The points $x_1^{(1)} $ and $x_2^{(1)}$ finding indicates the first iteration ending. In the second iteration these points $x_1^{(1)} $ and $x_2^{(1)}$ play the same role as the points~$x_1^{(0)} $ and $x_2^{(0)}$ in the first iteration and so on.

There are a number ways or the algorithm accuracy setting.
The simplest way of the iteration termination is stopping of the process by the following condition realization
\begin{equation}\label{stop}
    \left|\Delta x_1^{(n+1)}\right| + \left|\Delta x_2^{(n+1)}\right| \leq \epsilon,
\end{equation}
where $\epsilon$ is a some preassigned value. In particular, it can be precision of experimental data or any other quantity.

Let us consider some simple examples of the offered algorithm realization.

\section{Application of a method to some forms of the thermodynamic potential}
As examples, we consider the following model expressions for free energy:
\begin{enumerate}
    \item Free energy in a polynomial of the forth power form as a simplest example;
    \item Van der Waals free energy;
    \item Redlich-Kister form of free energy.
\end{enumerate}
Realization of the algorithm fulfilled in Mathematica 5.0 system, results output for graphical representation carried out in Excel system.

\subsection{A polynomial form of thermodynamic potential}
\begin{figure}
\includegraphics[width=5.3in]{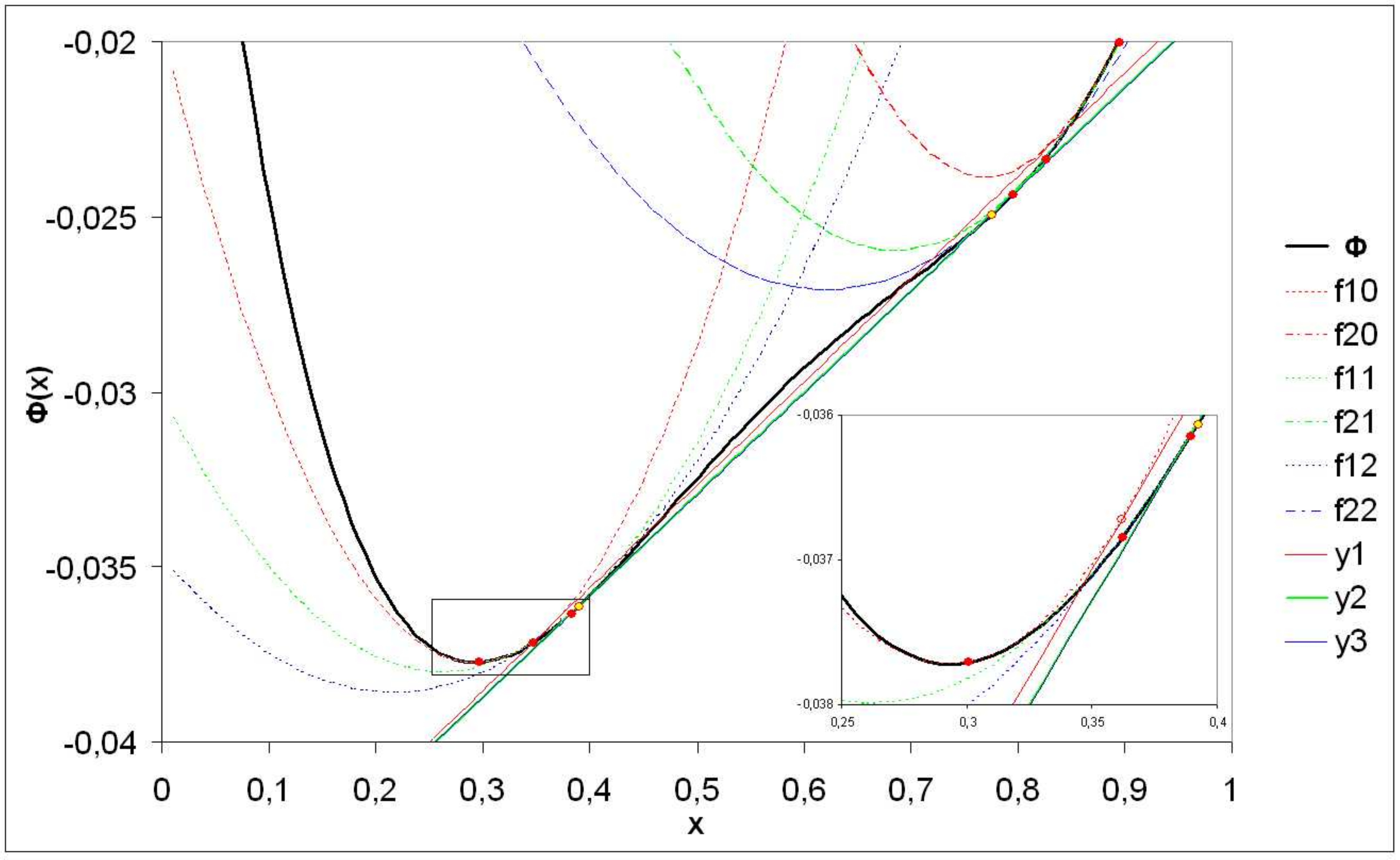} 
\caption{Search of the conode to model free energy in the form of the fourth power polynomial  $\Phi(x)=\frac{1}{2}x^4-\frac{20}{17}x^3+x^2-\frac{1}{3}x$. Notations: the black line is the function $\Phi(x)$ graph; colored dotted lines are graphs of the approximating parabolas in different approximations; straight lines are the tangents; the red points are the points $x_1^{(k)}$, $x_2^{(k)}$; the pricked red points are the points of contact between common tangents and approximating parabolas; the yellow points are the points of contact between conode and function $\Phi(x)$.}
\label{fig:pol}
\end{figure}

The simplest model potential having characteristic for a free energy in two-phase region behavior is the polynomial of the fourth power of $x$. This aim was achieved by the polynomial coefficients selection. Let us consider one of such polynomials
\begin{equation}
\label{st}
\Phi(x)=\frac{1}{2}x^4-\frac{20}{17}x^3+x^2-\frac{1}{3}x.
\end{equation}
Graph of this function is presented on figure~(\ref{fig:pol}).

The calculations show, that the relative error in determination of the tangency points after three steps does not exceed value $\frac{\left|x_i^{(3)} - \tilde x_i \right|}{\tilde x_i} < 5 \cdot 10^{-4}$, and after fourth step the error is less as $10^{-7}$  independent of the initial points~$x_1^{(0)}$, $x_2^{(0)}$ choice.

\subsection{The Van der Waals free energy}

\begin{figure}
\includegraphics[width=5.3in]{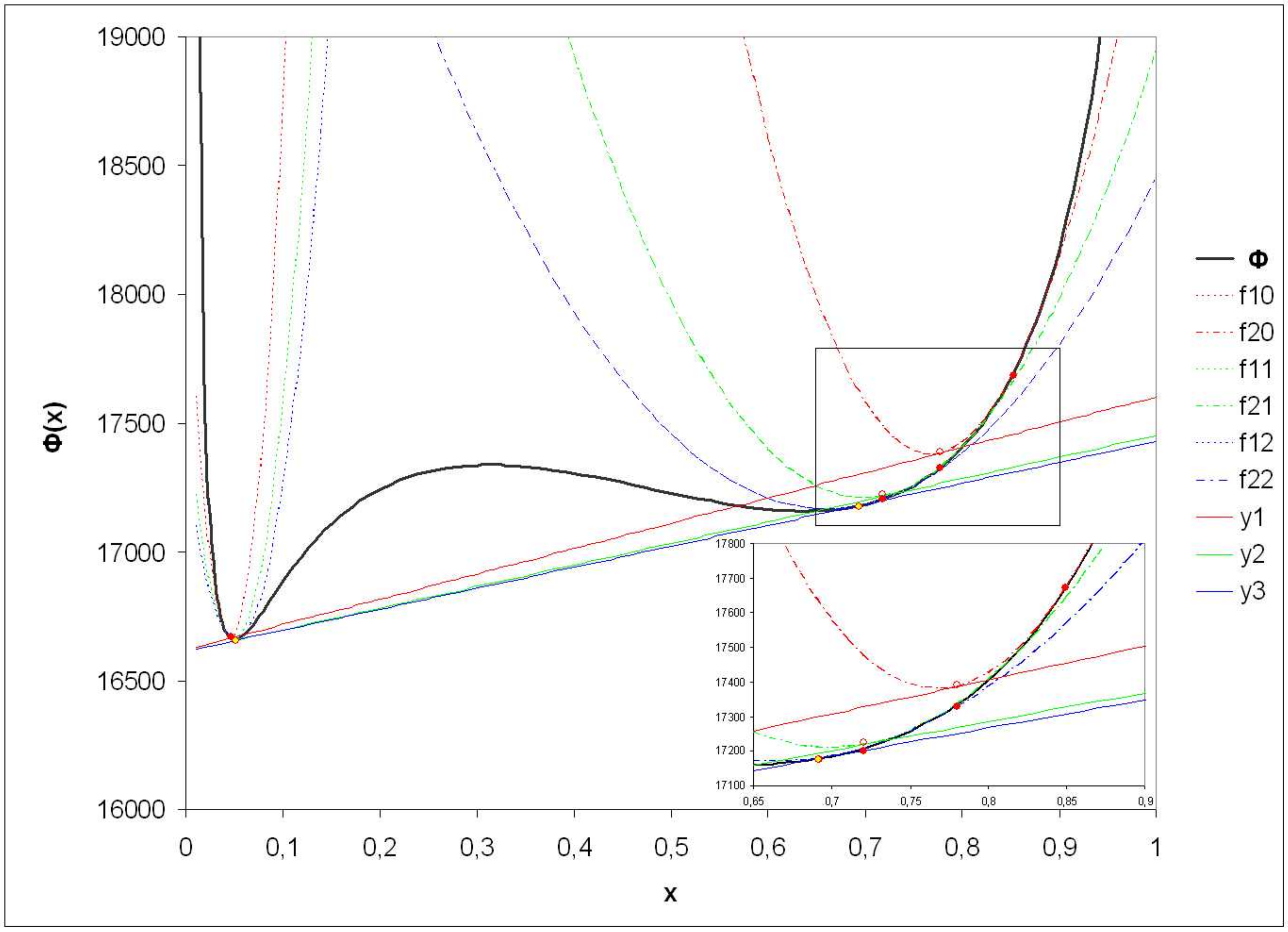} 
\caption{The calculations results for Van der Waals free energy $\Phi(x)=-2019.8\ln\left((1-x)/x\right)-8531x+85.34/x+21323$. The notations are the same as on figure~\ref{fig:pol}. }
\label{fig:vdw}
\end{figure}

The Van der Waals free energy has the following form:
\begin{equation}
\label{Gww}
G(P,T,N)=-R T N\ln(V-N b)-\frac{a N^2}{V}+P V+c(T,N),
\end{equation}
where $a,~b$ are the Van der Waals constants, $c(T,N)$ is an independent of the volume function. Let us introduce a new variable $x=Nb/V$:
\begin{equation}
\label{Gww1}
\begin{array}{r}
    {\displaystyle G(x)=-RTN\ln\left(\frac{Nb(1-x)}{x}\right)-\frac{aNx}{b}+\frac{bPN}{x}+c(T,N)   }\\
{\displaystyle  =\Phi(x)=-a_1\ln\left(\frac{1-x}{x}\right)-a_2x+\frac{a_3}{x}+a_4,  }
\end{array}
\end{equation}
where $a_i$  are the constants related with the Van der Waals parameters and variables $N,~T,~P$.
Results of the calculations for case $a_1=2019.8,~a_2=8531,~a_3=85.34,~ a_4=21323$ (these parameters correspond to carbon dioxide CO$_2$ at temperature $T\simeq 0.8\, T_c$ and pressure $P\simeq 20$~bar) presented on figure~(\ref{fig:vdw}).

\subsection{The Redlich-Kister potential}
The Redlich-Kister potential
\begin{equation}\label{redlich}
    \Phi=\sum_i x_i\Phi_i+RT\sum_i x_i\ln x_i+\sum_{j>i}x_ix_j L_{ij}\sum_{k=0}^n(x_i-x_j)^k
\end{equation}
is one of the most often used for modeling of the multicomponent systems  thermodynamic properties.

\begin{figure}
\includegraphics[width=5.3in]{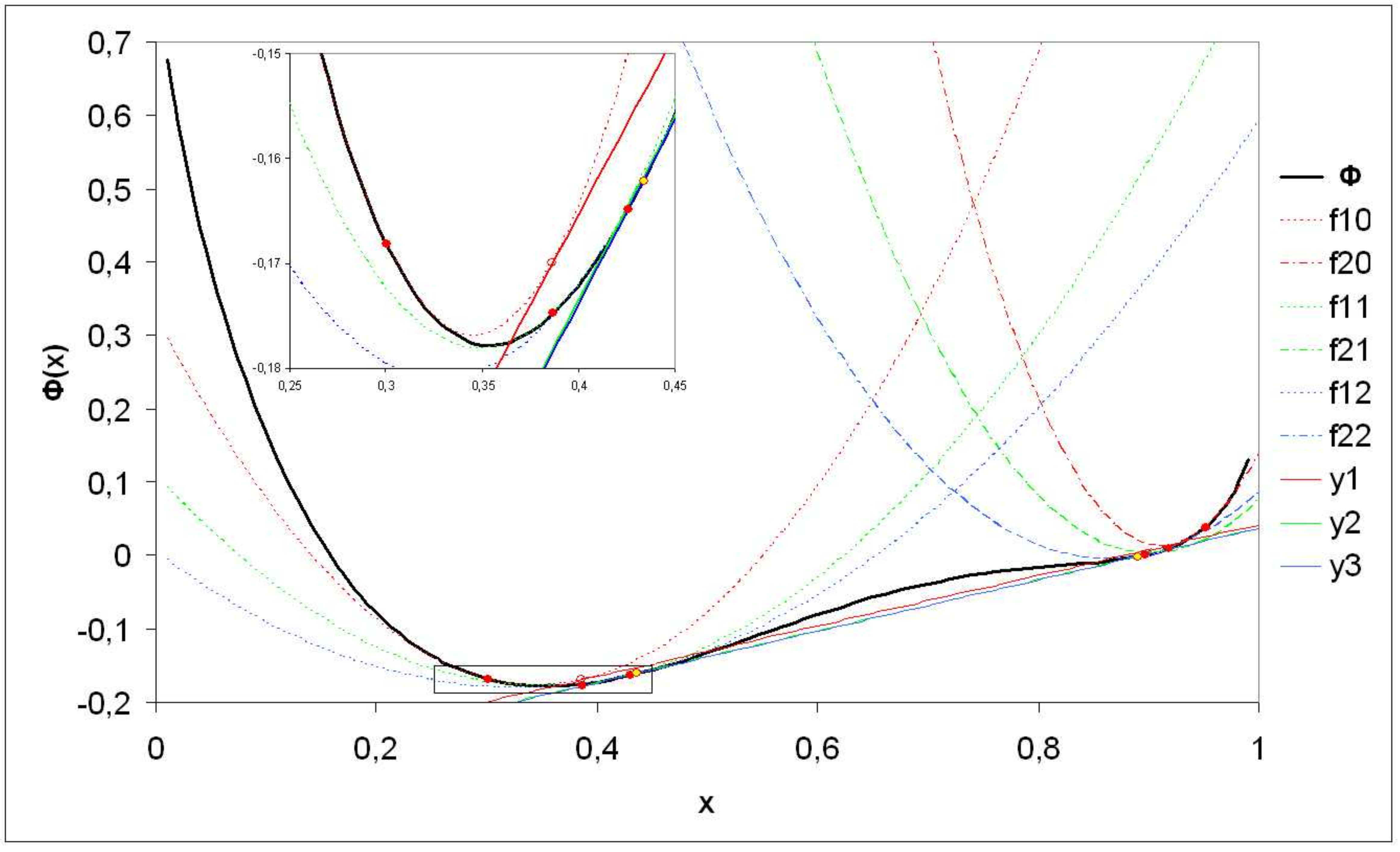} 
\caption{Search results of the conode for Redlich-Kister potential $\Phi\left(x \right)=2.5\left [x\ln x+(1-x)\ln(1-x)\right]+2.2x(1-x)\left(1+2x\right)+0.2x+0.8(1-x)$. The notations are the same as on figure~\ref{fig:pol}.}
\label{fig:rk}
\end{figure}

The calculations results for the Redlich-Kister potential 
\begin{equation}\label{red-kis}
    \begin{array}{r}
    {\displaystyle   \Phi(x)=2.5\left [x\ln x+(1-x)\ln(1-x)\right] }\\  \\
{\displaystyle +2.2x(1-x)\left(1+2x\right)+0.2x+0.8(1-x)   }
\end{array}
\end{equation}
are presented on figure~(\ref{fig:rk}).

\section{Conclusion}

The presented method for determination of coexisting phases compositions qualitatively differs from usually used methods based on a free energy numerical minimization~\cite{Hillert2,W,Udovsky,Em,Ch,CZG,Sim}.

Using this method to several model thermodynamic potentials has shown the efficiency of the algorithm. As a rule, it is enough to three iteration to obtain accuracy of the coexisting phases compositions up to $10^{-4}$. Each of the iterations includes nine computing operations (evaluation function and its first and second derivatives in the initial points $x_1^{(0)},\,  x_2^{(0)}$, and then  $\Delta x^{(n+1)}$, $x_1^{(n+1)}$ и $x_2^{(n+1)}$). The rigorous analytical estimate of degree of convergence will be derived later.

The authors gratefully acknowledge the financial support from Program of Russian Ministry of Education and Science ``Scientific and pedagogical personnel of innovative Russia'' on 2009--2013, Project of Russian Ministry of Education and Science-Section 2.1.2-Fundamental researches in technical sciences (Project No.11324), and Department of Chemistry and Material Science of Russian Academy of Sciences. We are grateful also to Prof.~A.L.~Udovsky for useful discussion of the paper.

\end{document}